\documentclass[epjCONF,columns]{svjour} 
\usepackage{graphics}
\usepackage[varg]{txfonts} 
\usepackage[latin1]{inputenc}
\session-title{HadronColliderPhysics Symposium 2011}
\begin{document}
\title{Search for $B_{s}\rightarrow\mu^{+}\mu^{-}$ and $B_{0}\rightarrow\mu^{+}\mu^{-}$ decays in CMS}
\author{Luca Martini\inst{1}\fnmsep\thanks{\email{lmartini@cern.ch}} for the CMS collaboration}
\institute{$^{1}$Univ. Siena \& INFN Pisa}
\abstract{
A search for the rare decays $B_{s}\rightarrow\mu^{+}\mu^{-}$ and $B_{0}\rightarrow\mu^{+}\mu^{-}$ performed in $pp$ collisions at $\sqrt{s} = 7$~TeV is presented. The data sample, collected by the CMS experiment at the LHC, corresponds to an integrated luminosity of 1.14~fb$^{-1}$. In both cases the number of events observed after all selection requirements is consistent with expectations from background and standard model signal predictions. The resulting upper limits on the branching fractions are $BF(B_{s}\rightarrow\mu^{+}\mu^{-}) < 1.9\times 10^{-8}$  and $BF(B_{0}\rightarrow\mu^{+}\mu^{-}) < 4.6\times 10^{-9}$ at 95\% confidence level (CL). Furthermore, the combination of the results of the search for the decay  $B_{s}\rightarrow\mu^{+}\mu^{-}$ by the CMS and LHCb experiments is presented. The combined upper limit is $BF < 1.1\times 10^{-8}$ at 95\%~CL.
} 
\maketitle
\section{Introduction}
\label{intro}
\begin{figure}
\resizebox{1.\columnwidth}{!}{
  \includegraphics{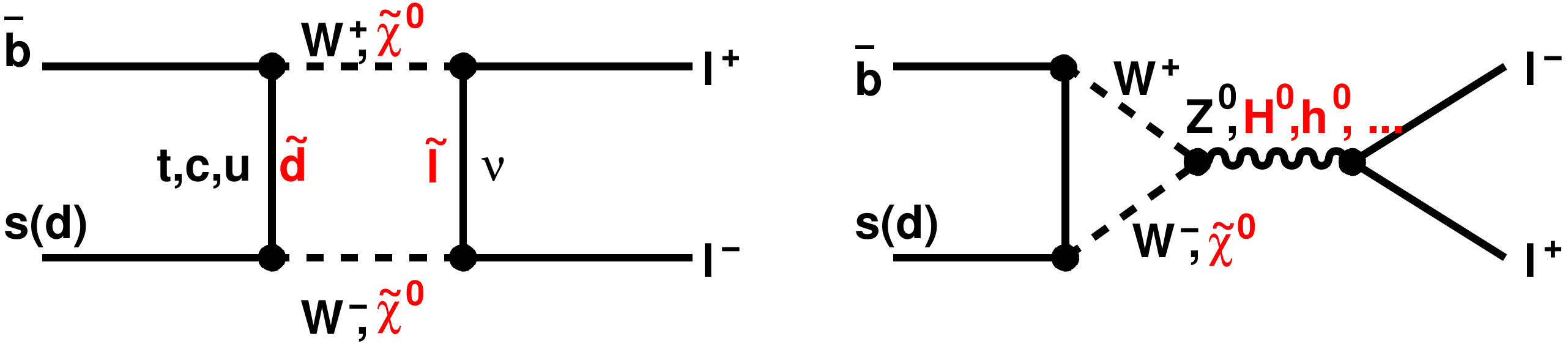}
   }
\caption{Illustration of the rare decays $B_{s(d)}^0 \rightarrow \mu^+ \mu^-$.  
In the SM these decays proceed through  box (left) and penguin (right) interactions.}
\label{fig:diagr} 
\end{figure}

In the standard model (SM) of particle physics, flavor changing neutral current decays are highly suppressed (Fig.~\ref{fig:diagr}):
they are forbidden at tree level and can only proceed through higher-order loop diagrams, 
they are helicity suppressed by factors of $(m_l/m_B)^2$, where $m_l$ and $m_B$ are the masses of the lepton and $B$ meson 
and they require an internal quark annihilation within the $B$ meson.
The SM predictions (Tab.~\ref{tab:SM_BF}) are significantly enhanced in several extensions of the SM, although in some cases the decay rates are lowered.
A (blind) search for the rare decays $B_{s}\rightarrow\mu^{+}\mu^{-}$ and $B_{0}\rightarrow\mu^{+}\mu^{-}$ is presented here, using $1.14$~fb$^{-1}$ of integrated luminosity collected by the CMS experiment in 2011 \cite{BsAndBd}. 
\begin{table}
\centering
\caption{Branching fraction SM predictions for $B_{s}\rightarrow\mu^{+}\mu^{-}$ and $B_{0}\rightarrow\mu^{+}\mu^{-}$.}
\label{tab:SM_BF} 
\begin{tabular}{ll}
\hline\noalign{\smallskip}
Decay channel & BF SM predictions \\
\noalign{\smallskip}\hline\noalign{\smallskip}
$B_{s}\rightarrow\mu^{+}\mu^{-}$ & $(3.2\pm 0.2)\times 10^{-9}$ \\
$B_{0}\rightarrow\mu^{+}\mu^{-}$ & $(1.1\pm 0.1)\times 10^{-10}$ \\
\noalign{\smallskip}\hline
\end{tabular}
\end{table}
\section{A blind analysis}
We perform a counting experiment in the dimuon mass distribution centered on the $B_s (B_0)$ meson mass (see Tab.~\ref{tab:windows} for the region definitions).
\begin{table}
\centering
\caption{Region definitions}
\label{tab:windows} 
\begin{tabular}{ll}
\hline\noalign{\smallskip}
Region definition & Invariant mass (GeV) \\
\noalign{\smallskip}\hline\noalign{\smallskip}
overall window & $4.90<m_{\mu_1\mu_2}<5.90$ \\ 
blinding window & $5.20<m_{\mu_1\mu_2}<5.45$ \\ 
$B_{s}\rightarrow\mu^{+}\mu^{-}$  window & $5.20<m_{\mu_1\mu_2}<5.30$ \\ 
$B_{0}\rightarrow\mu^{+}\mu^{-}$ window & $5.30<m_{\mu_1\mu_2}<5.45$ \\ 
\noalign{\smallskip}\hline
\end{tabular}
\end{table}
The background is estimated from the sidebands and from Monte Carlo (MC) simulation.

The analysis uses a relative normalization to the well-measured decay $B^\pm \rightarrow J/\psi (\mu^+\mu^-) K^\pm$ to avoid a dependence on the uncertainties of the $b\bar{b}$ production cross section and luminosity measurements. The  $B_{s}\rightarrow\mu^{+}\mu^{-}$ branching fraction is measured using
\begin{equation}
BF(B_{s}\rightarrow\mu^+\mu^-)=\frac{N_S}{N_{obs}^{B^\pm}}\frac{f_u}{f_s}\frac{\epsilon^{B^\pm}_{tot}}{\epsilon_{tot}}BF(B^\pm\rightarrow J/\psi K^\pm)
\end{equation}
where $BF(B^\pm\rightarrow J/\psi K^\pm)$ is the branching fraction $B^\pm\rightarrow J/\psi K^\pm$; $N_S/N_{obs}$ is the background-subtracted number of observed signal candidates in the signal window over the number of reconstructed $B^\pm$ events; $\epsilon^{B^\pm}_{tot}/\epsilon_{tot}$ is the ratio of the signal total efficiency over the $B^\pm$ efficiency and $f_u/f_s$ is the ratio of the $B^\pm$ and $B_s$ meson production fractions.

Since the mass resolution in the CMS detector depends strongly on the pseudorapidity ($\eta$) of the reconstructed particles, the analysis is performed in two "channels": barrel (if both muons have $|\eta| < 1.4$) and endcap (elsewhere), that are then combined for the final result. 
The determination of the signal efficiency in this analysis depends on MC simulation, that is validated through two samples of fully reconstructed $B$ decays: 
(1) the decay $B^\pm \rightarrow J/\psi K^\pm$, that provides a high-statistics sample to allow fine-grained comparisons;
(2) the decay $B_s \rightarrow J/\psi \phi$ that is essential to compare $B_s$ mesons in data and MC simulations and to estimate systematic uncertainties for the analysis efficiency.

\subsection{The signal channel $B_{s(d)}\rightarrow\mu^+\mu^-$}
\begin{figure}
\resizebox{1.\columnwidth}{!}{
  \includegraphics{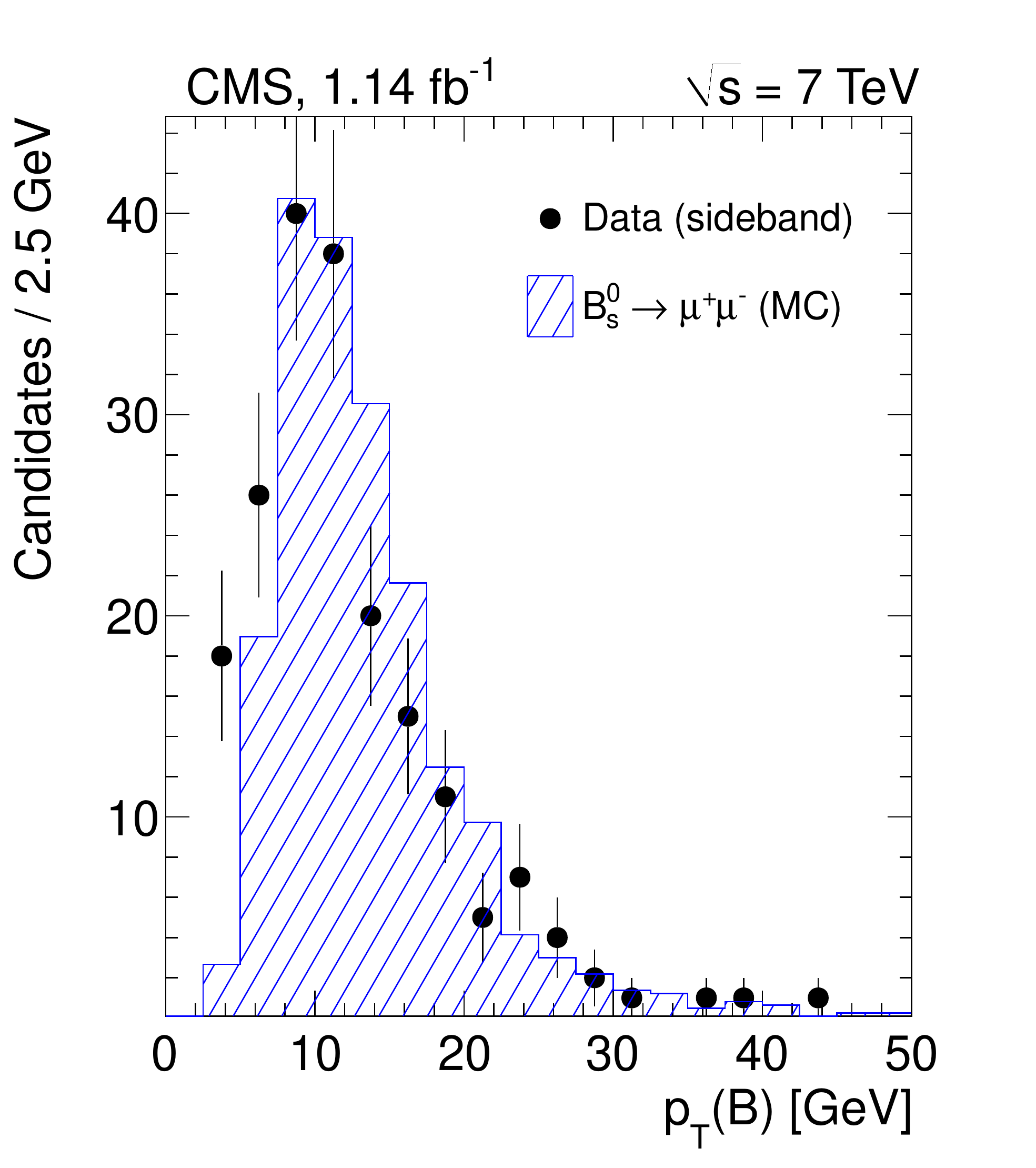}
  \includegraphics{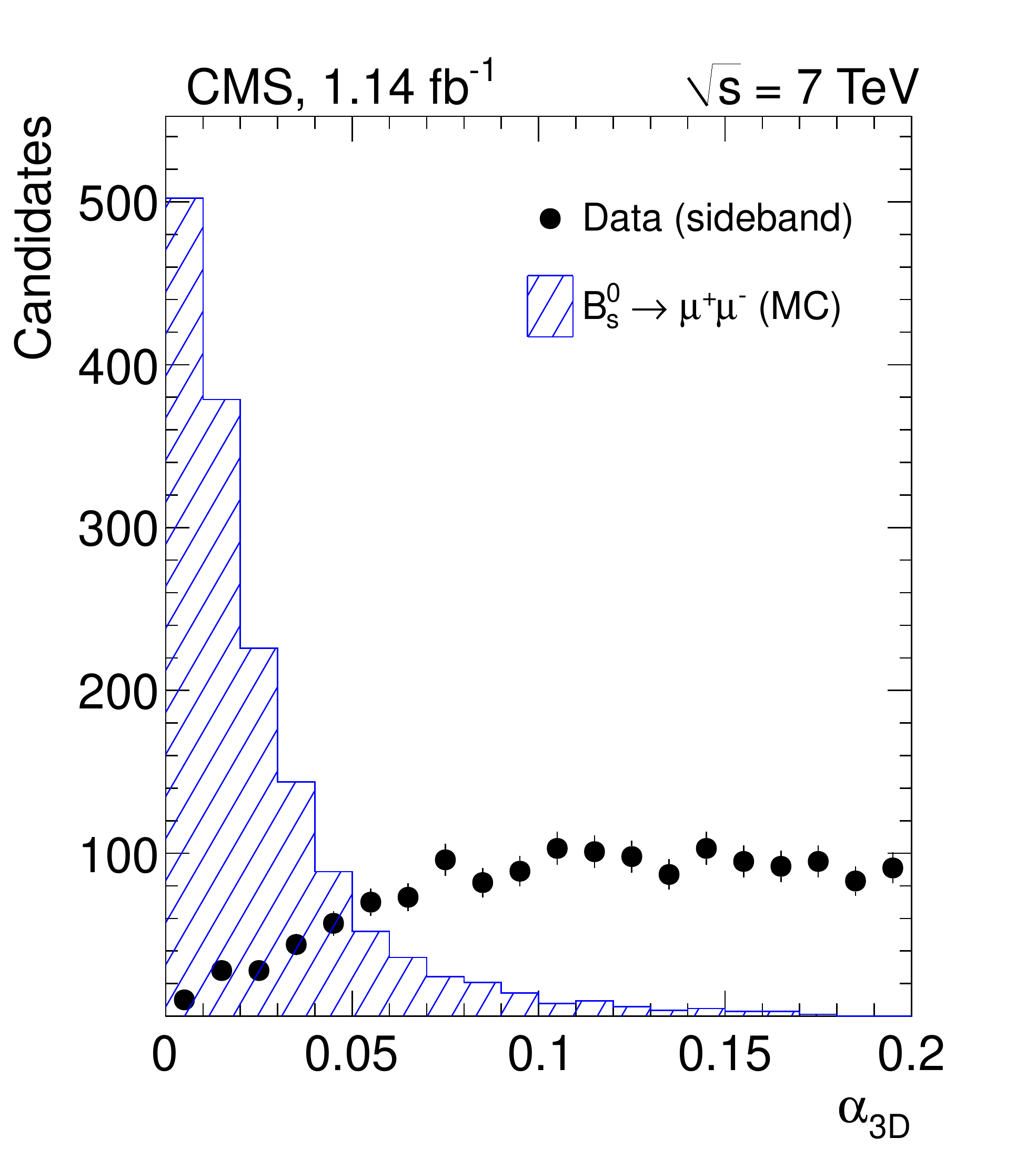}
   }
   \resizebox{1.\columnwidth}{!}{
   \includegraphics{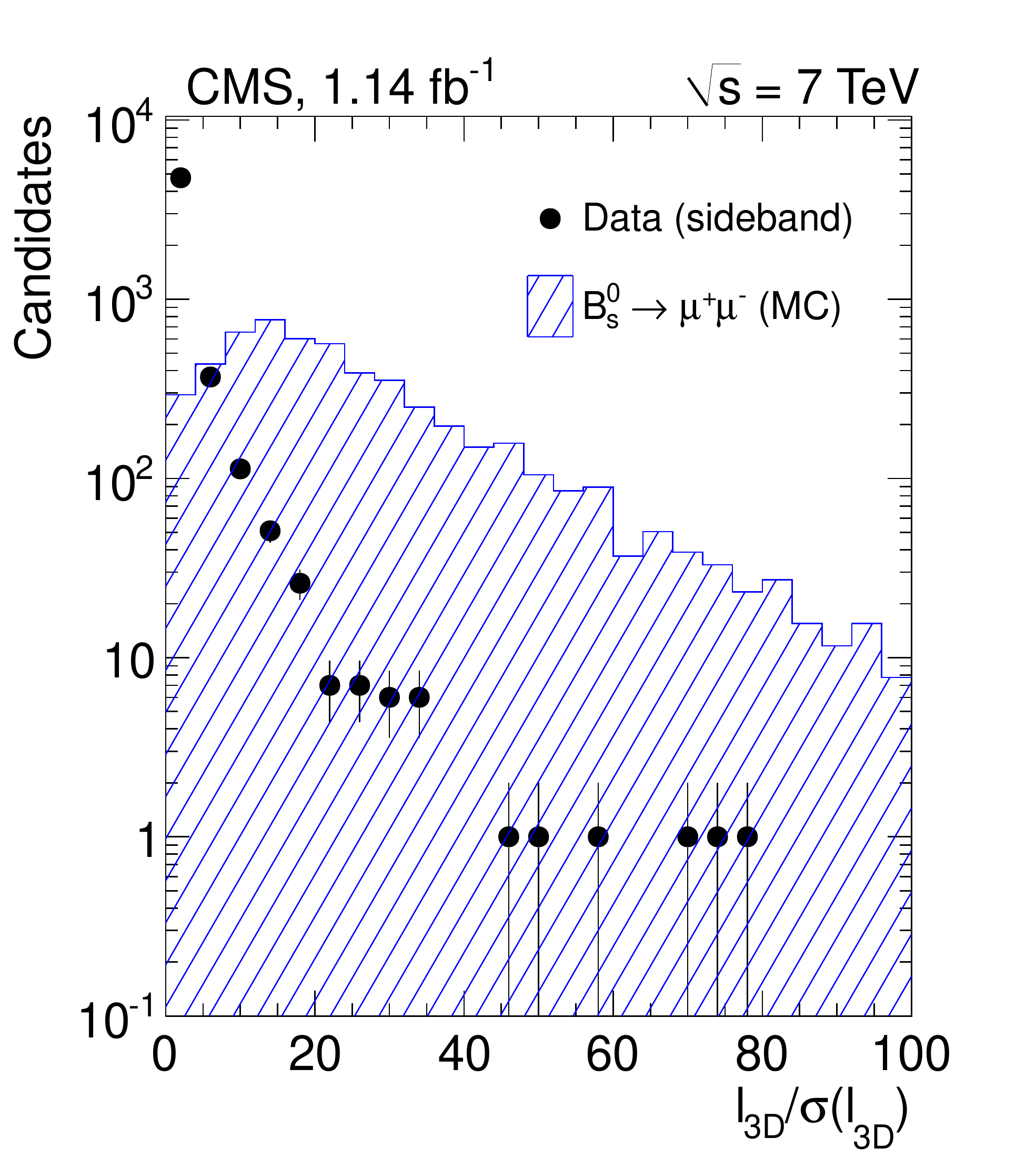}
  \includegraphics{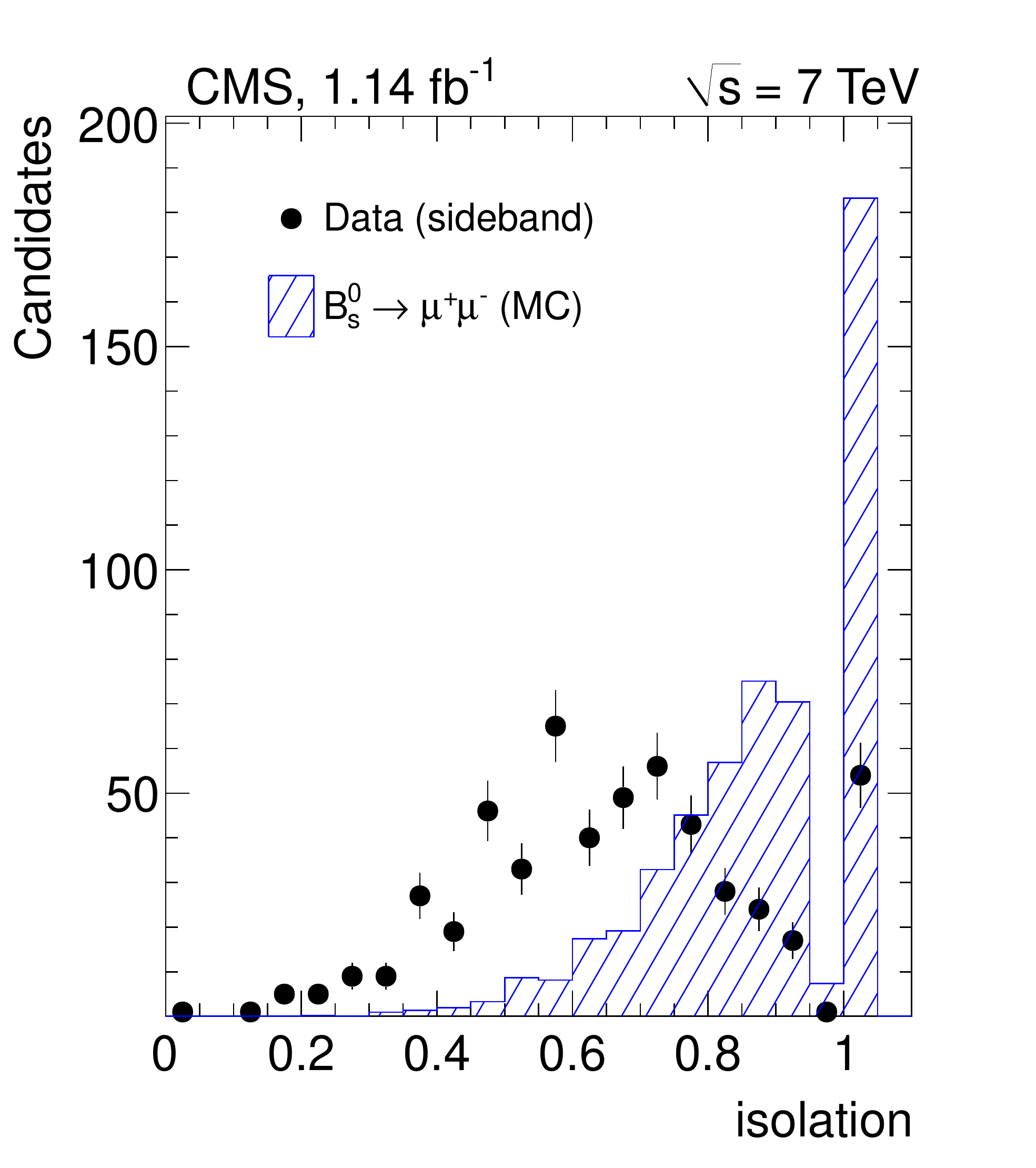}
   }
\caption{Comparison of MC signal and sideband data distributions, for the transverse momentum (top left), the 3D pointing angle (top right), the flight length significance (bottom left), and the isolation (bottom right). The MC histograms are normalized to the number of events in the data.
}
\label{fig:sgDistro} 
\end{figure}

The events are selected with a two-level trigger system. The first hardware level only requires two muon candidates, while the high-level software trigger uses additional information from the silicon tracker.
For the offline event selection, variables related to the muons, the primary vertex, and the $B_s$ candidate with its associated secondary vertex are calculated. 
The candidate's secondary vertex and its momentum are used to select a matching primary vertex based on the distance of closest approach.
The flight length significance is computed as the ratio of the flight length to its error: $S_{3D} = l_{3d}/\sigma(l_{3d})$. 
The pointing angle $\alpha$ is defined as the angle in three dimensions between the B candidate momentum and the vector from the primary vertex to the B decay vertex.
The isolation $I$ is determined from the B candidate transverse momentum and other charged tracks in a cone with radius $\Delta R = \sqrt{(\Delta\eta)^2+(\Delta\phi)^2} =1$  around the B momentum as $I=\frac{p_T(B)}{p_T(B)+\sum_{trk}p_T}$.
A comparison of several distributions for signal MC and for sideband background data events is shown in Fig.~\ref{fig:sgDistro}.

\subsection{Background}
The background sources that mimic the signal topology can be grouped into three categories:
(1) $qq$ events (where $q = b$, $c$) with $q \rightarrow X \mu \bar{\nu}$ (prompt or cascade) decays of both $q$ hadrons.  
(2) Events where a true muon is combined with a hadron misidentified as a muon (punch-through or in-flight decay of a hadron).
(3) Rare $B^0$, $B^\pm$, $B_s^0$ and $\Lambda_b$ decays, mostly from semileptonic decays: 
peaking backgrounds from rare decays, where a heavy particle decays into a pair of hadrons (examples include $B_s^0 \rightarrow K^+K^-$, $\Lambda_b \rightarrow p K^-$) and non-peaking backgrounds from rare $B^0$, $B^\pm$ and $B_s^0$ decays, comprising hadronic, semileptonic, and radiative decays.

\subsection{The normalization and control channels}
The reconstruction of $B^\pm \rightarrow J/\psi K^\pm$ candidates starts from two opposite-charged muons with invariant mass $3.0 < m_{\mu_1\mu_2} < 3.2$~GeV, which are combined with a track, assumed to be a kaon, fulfilling $p_T > 0.5$~GeV. The distance of closest approach between all pairs among the three tracks is required to be less than 1~mm. All three tracks are used in the vertex reconstruction. 

The $B_s\rightarrow J/\psi \phi$ decay channel is used to check that exclusive $B_s$ meson decays are correctly simulated in MC. The reconstruction candidates starts from two opposite-charged muons, which are combined with two tracks assumed to be kaons, fulfilling $p_T > 0.5$~GeV and $|\eta| < 2.4$. The distance of  closest approach between all pairs among the four tracks is required to be less than 1~mm. All four tracks are used in secondary vertex reconstruction. The two muons of the candidate must have an invariant mass $3.0 < m_{\mu_1\mu_2} < 3.2$~GeV. The two kaons must have an invariant mass of $0.995 < m_{KK} < 1.045$~GeV and $\Delta R < 0.25$.

\begin{figure}
\resizebox{1.\columnwidth}{!}{
  \includegraphics{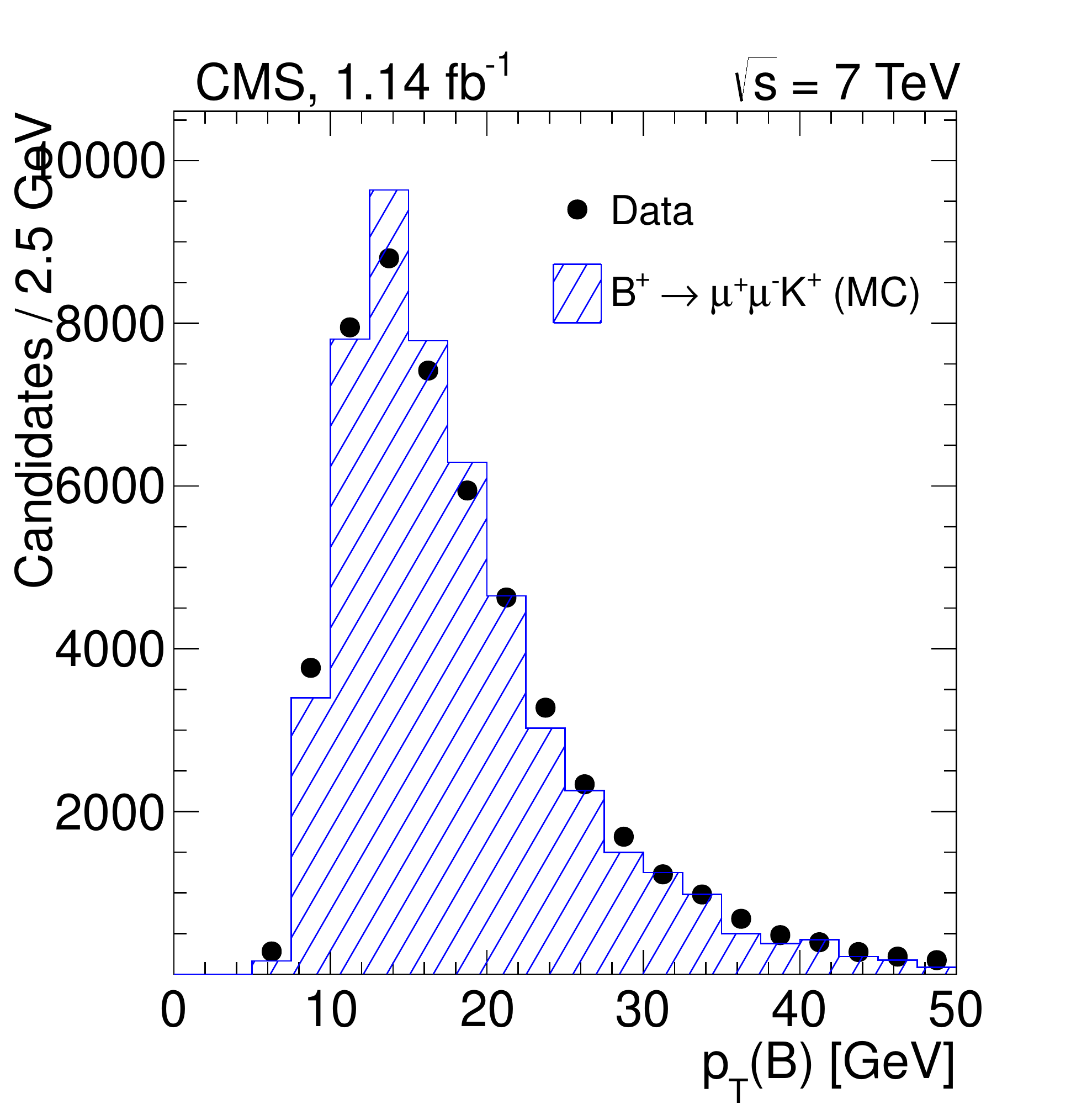}
  \includegraphics{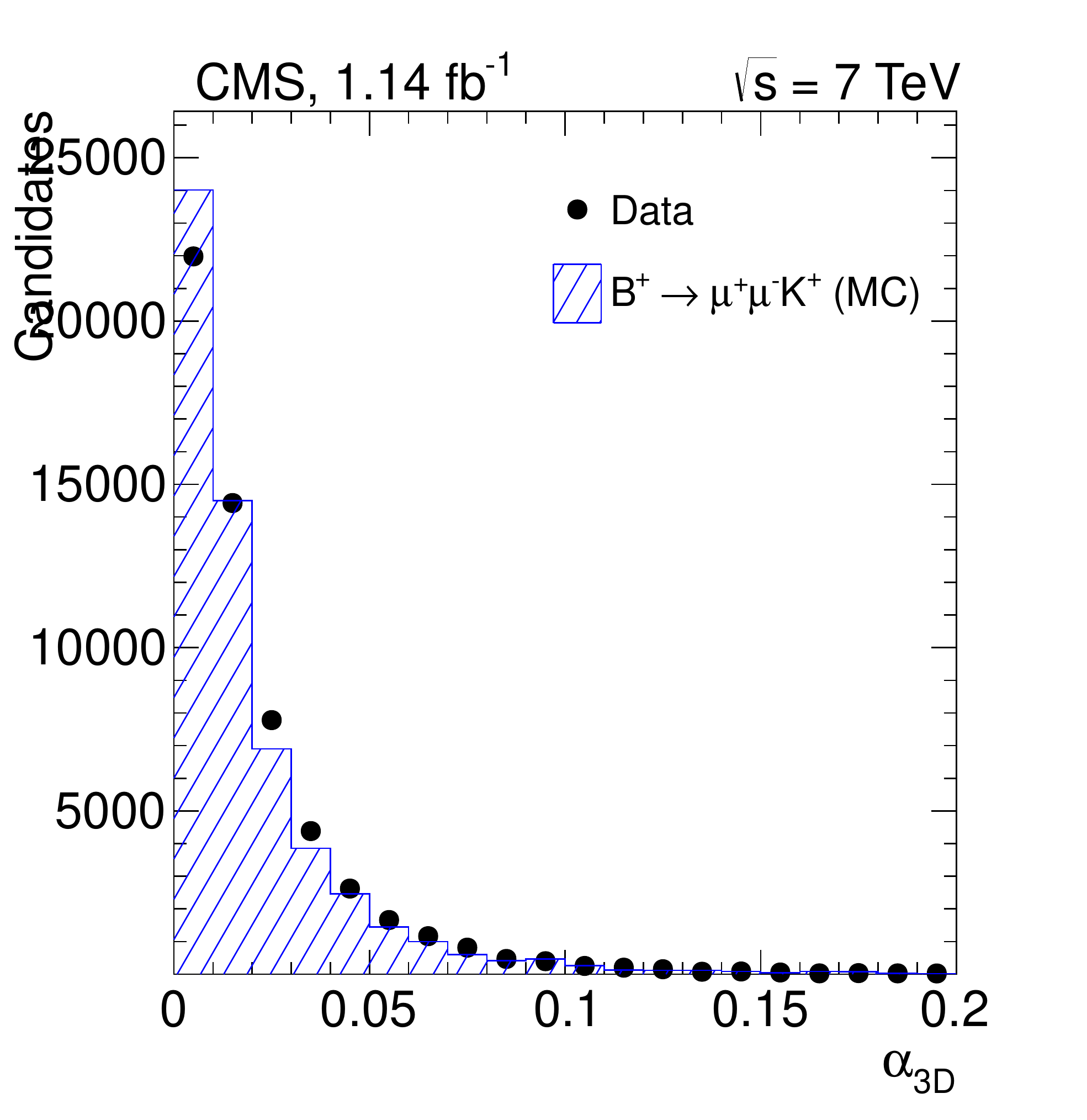}
   }
   \resizebox{1.\columnwidth}{!}{
   \includegraphics{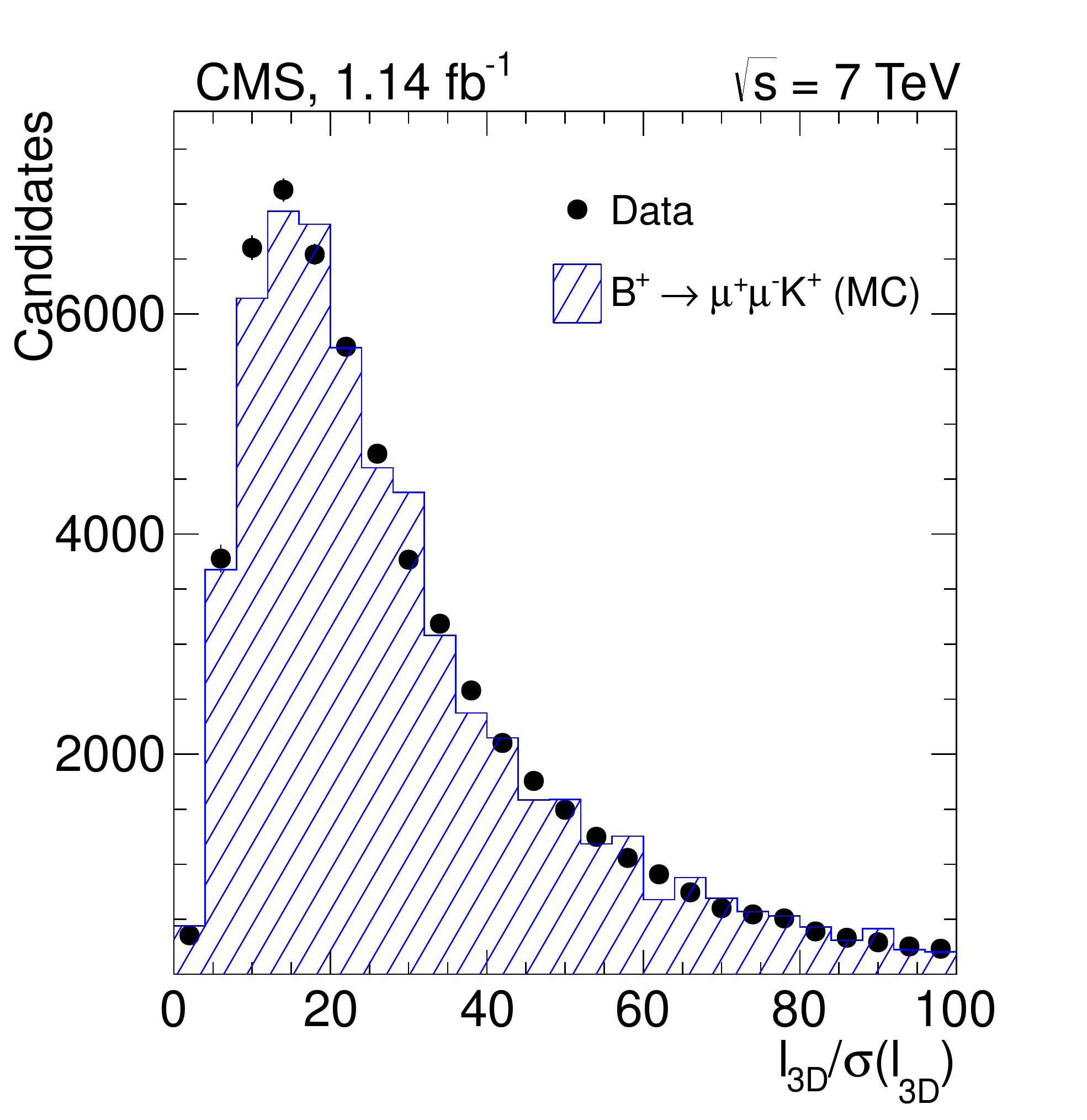}
  \includegraphics{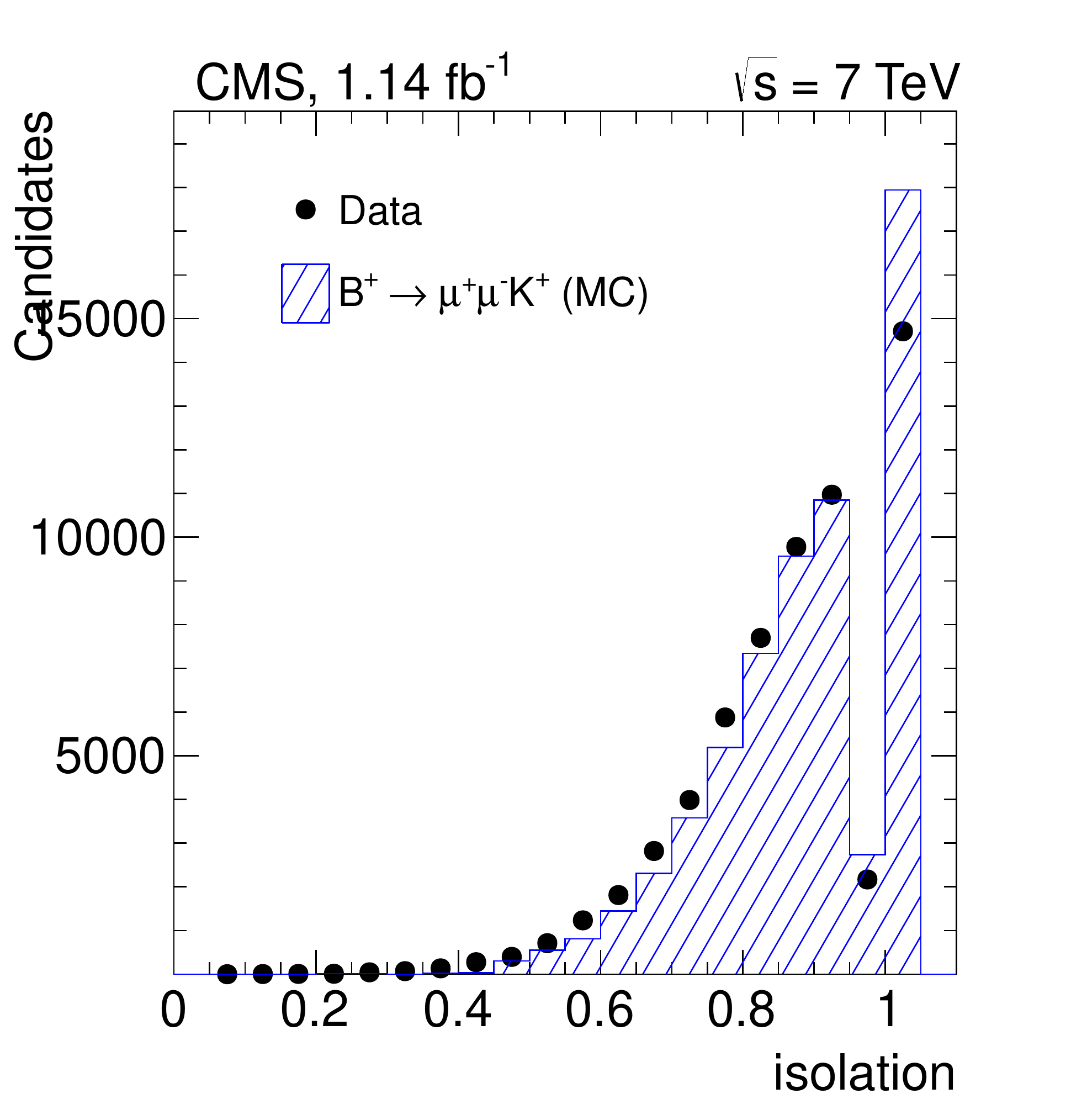}
   }
\caption{Comparison of measured and simulated $B\pm\rightarrow J/\psi K^\pm$ distributions, for the transverse momentum (top left), the 3D pointing angle (top right), the flight length significance (bottom left), and the isolation (bottom right). The MC histograms are normalized to the number of events in the data.
}
\label{fig:noDistro} 
\end{figure}

\begin{figure}
\resizebox{1.\columnwidth}{!}{
  \includegraphics{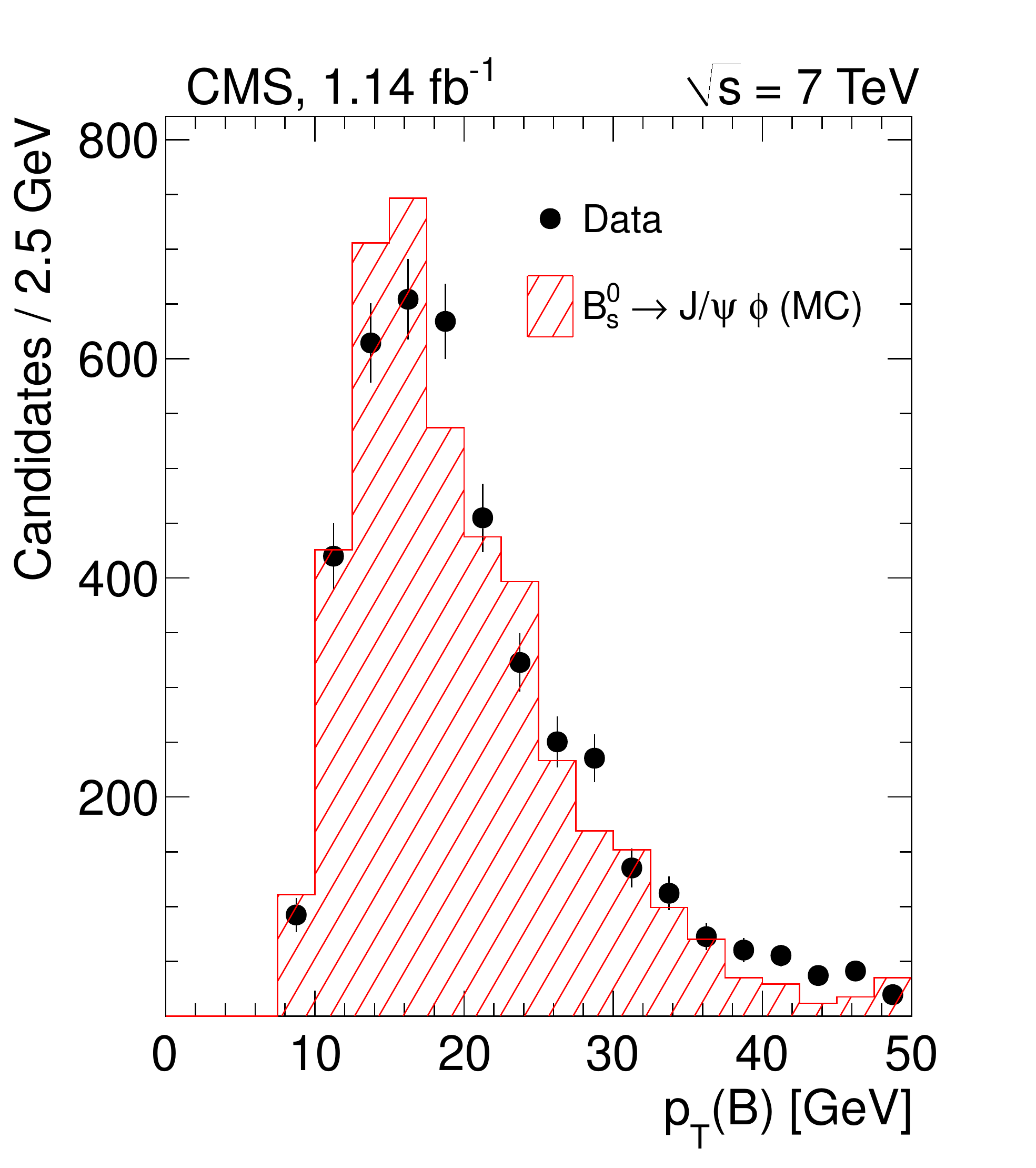}
  \includegraphics{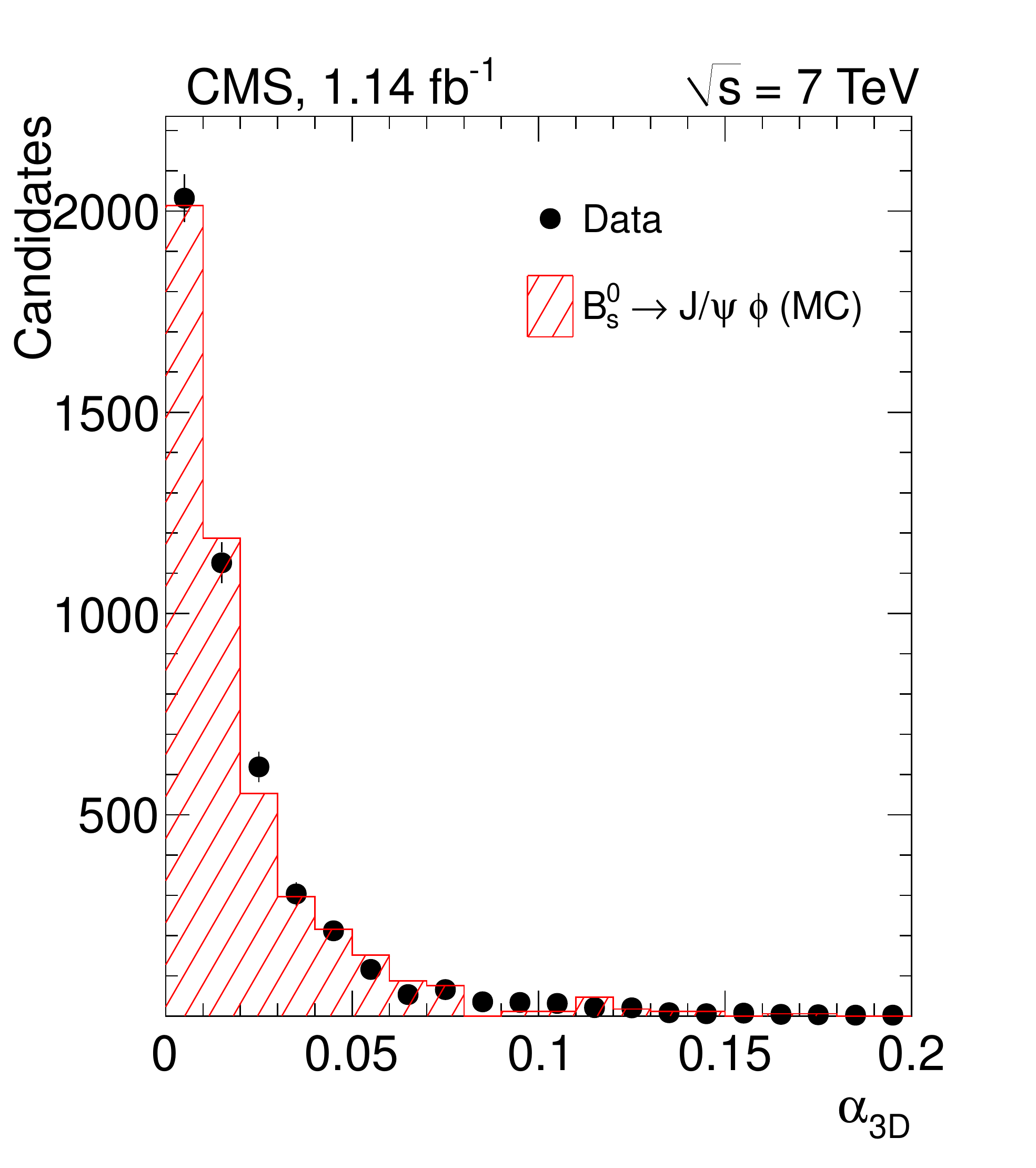}
   }
   \resizebox{1.\columnwidth}{!}{
   \includegraphics{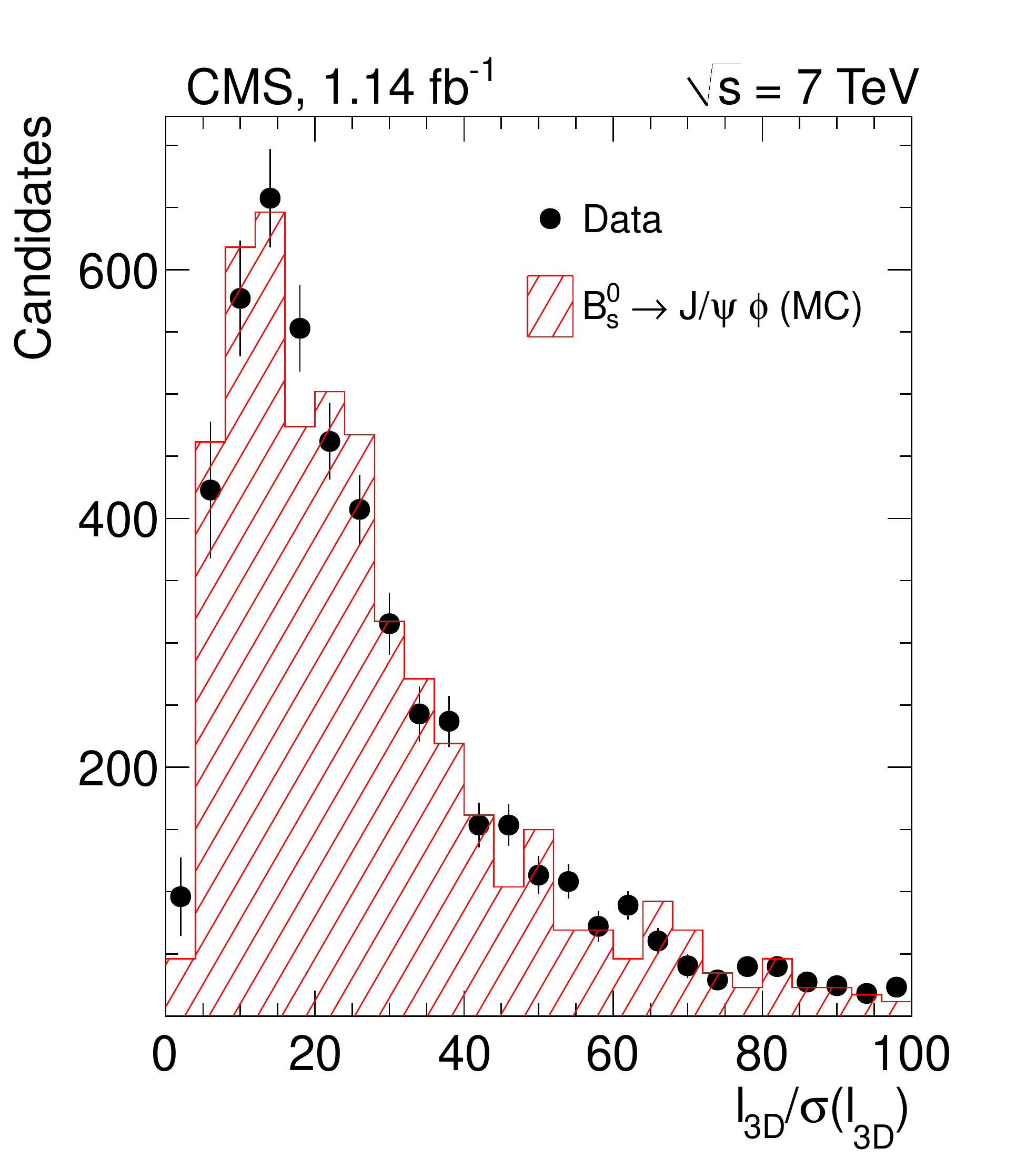}
  \includegraphics{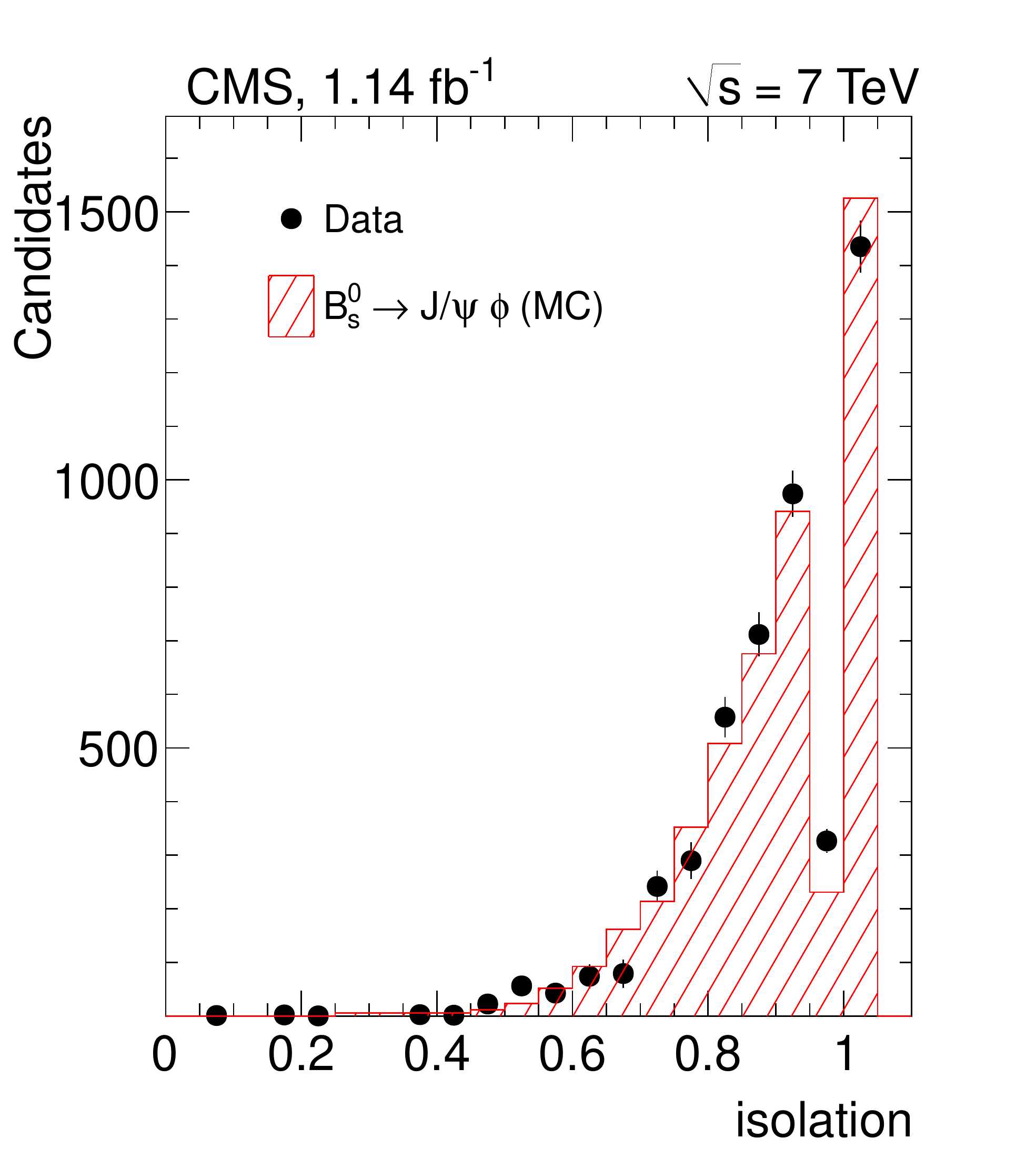}
   }
\caption{Comparison of measured and simulated $B_0\rightarrow J/\psi\phi$ distributions, for the transverse momentum (top left), the 3D pointing angle (top right), the flight length significance (bottom left), and the isolation (bottom right). The MC histograms are normalized to the number of events in the data.
}
\label{fig:csDistro} 
\end{figure}

Comparisons of several distributions for the normalization and control candidates, between MC simulation and sideband-subtracted data, are shown in  Fig.~\ref{fig:noDistro} and Fig.~\ref{fig:csDistro}.

\subsection{Main systematic uncertainties}
The main source of systematic uncertainty for the branching fractions and the production mechanisms is the error of $b$-quark hadronization fractions $f_u$ and $f_s$ and it is equal to 12.5\%. 
The acceptance dependence on different production mechanisms has been evaluated for signal and normalization samples and the difference results in 4\%. 
The uncertainty on the analysis efficiency is determined as the quadratic sum of all the efficiency differences between data and MC simulation in the control sample (7.9\%). 
For the background estimation, candidate counts from the sidebands are interpolated in the signal boxes to estimate the background candidate yield. This results in a difference of 4\%. 
For the peaking rare B decays, a set of rare hadronic backgrounds was generated, simulated, and passed through the full detector simulation and evaluated taking into account the CMS muon misidentification.

\section{Results}
To optimize the analysis selection, a grid of selection requirements was defined and searched for the best expected upper limit (Tab.~\ref{tab:opt_cuts}).

\begin{table}
\centering
\caption{Optimized selections for the best upper limit.}
\label{tab:opt_cuts} 
\begin{tabular}{llll}
\hline\noalign{\smallskip}
Selection & Barrel & Endcap & Unit  \\
\noalign{\smallskip}\hline\noalign{\smallskip}
$p_{T,\mu_1} > $ & 4.5 & 4.5 & GeV \\
$p_{T,\mu_2} > $ & 4.0 & 4.0 & GeV  \\
$p_{T,B} > $ & 6.5 & 6.5 & GeV \\
$I > $ & 0.75 & 0.75 &  \\
$\chi/dof < $ & 1.6 & 1.6 & \\
$\alpha < $ & 0.050 & 0.025 & rad \\
$S_{3D} > $ & 15.0 & 20.0 & \\
$d_{ca}^0 > $ & n/a & 0.015 & cm \\
\noalign{\smallskip}\hline
\end{tabular}
\end{table}

\begin{figure}
\resizebox{1.\columnwidth}{!}{
  \includegraphics{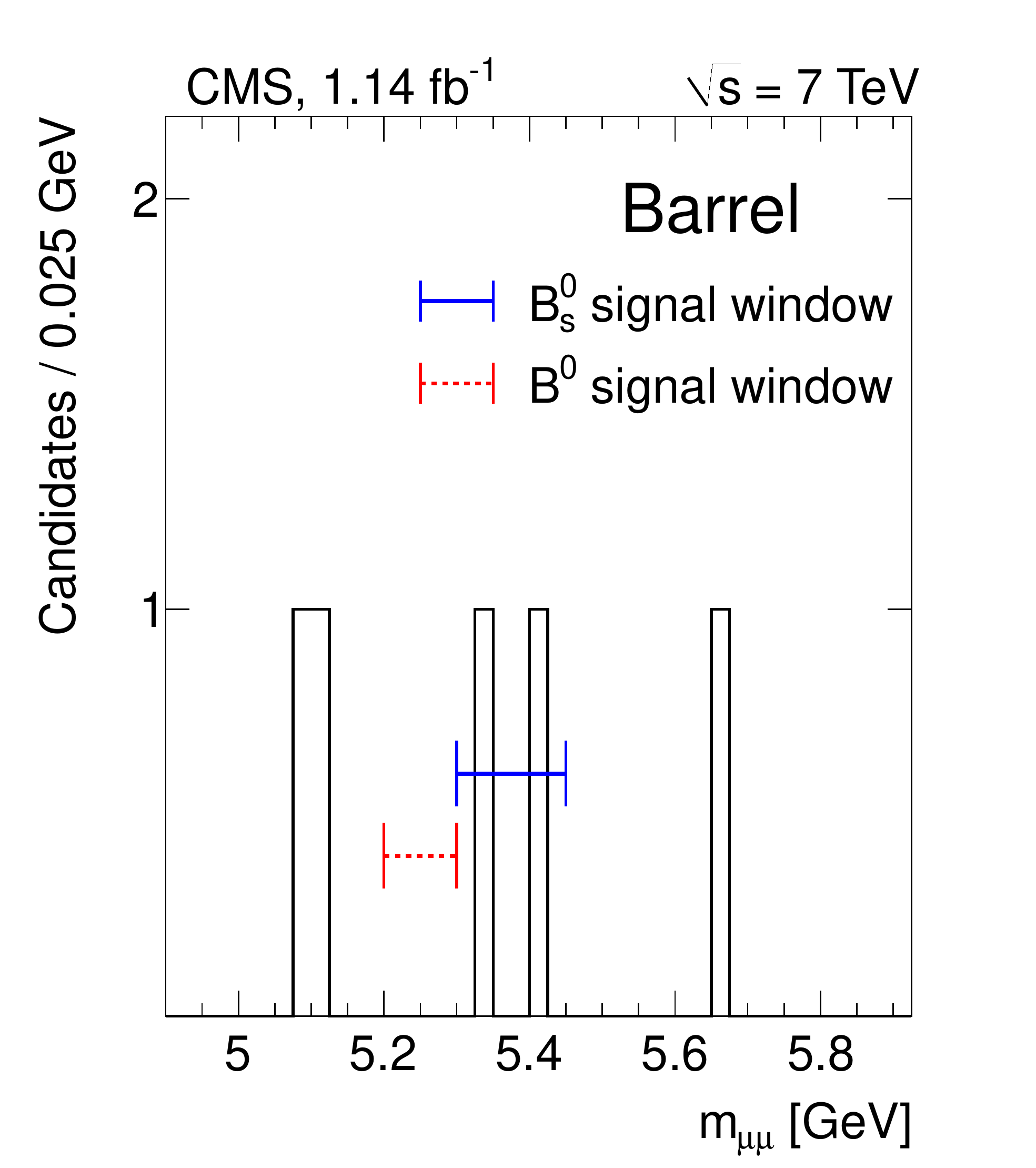}
  \includegraphics{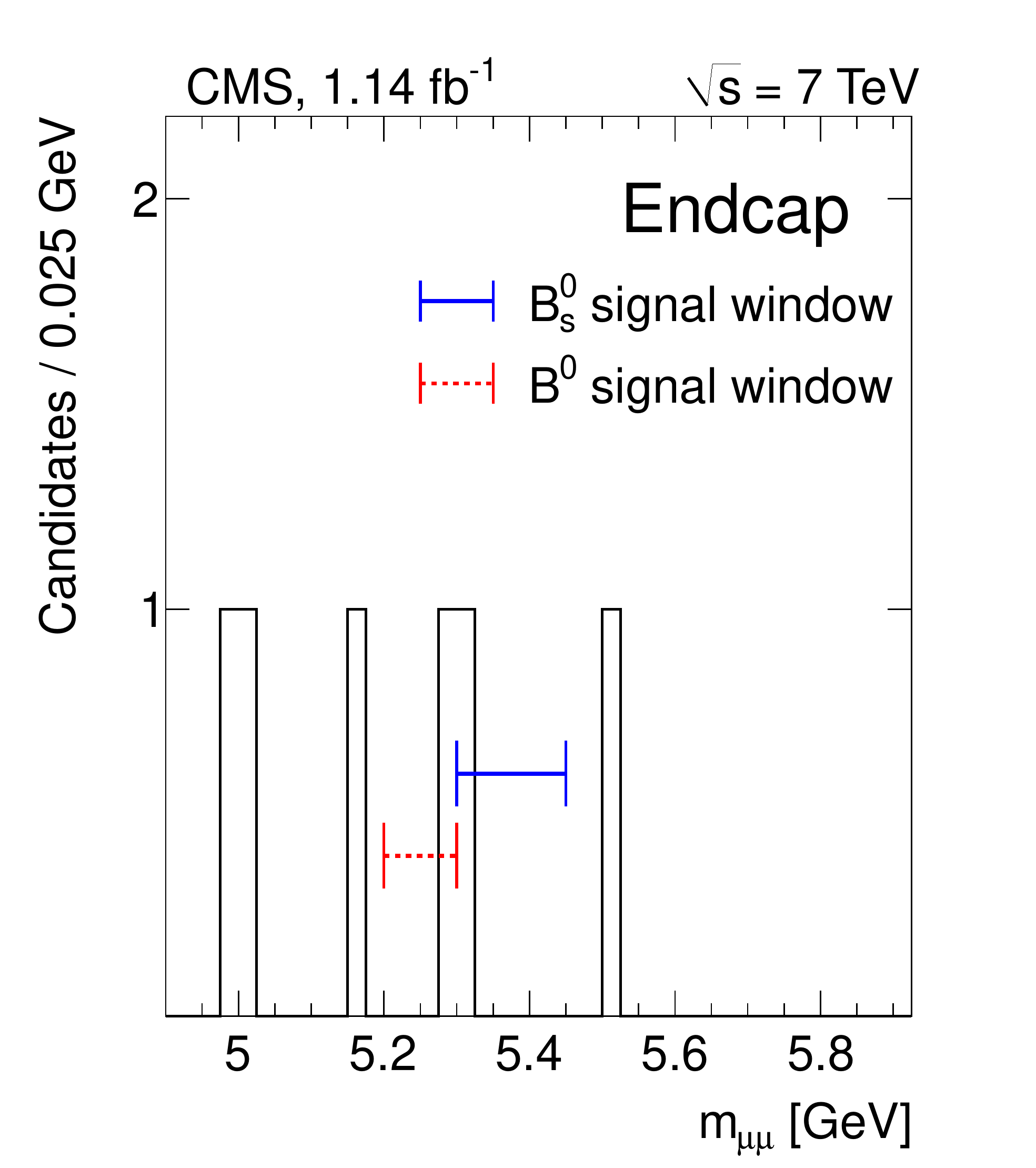}
   }
\caption{Unblinded dimuon invariant mass distributions in the barrel (left) and the endcap (right) channels.
}
\label{fig:invMass} 
\end{figure}

Tab.~\ref{tab:finalTab} summarizes all numbers relevant for the extraction of the upper limit: the final efficiencies, the number of observed background events in the sidebands, the expected number of resonant and non-resonant background events in the signal windows and the number of observed events. Fig.~\ref{fig:invMass} shows the "unblinded" invariant mass distributions.
\begin{table}
\centering
\caption{Final results for each signal "channel" used for the extraction of the upper limit.}
\label{tab:finalTab} 
\begin{tabular}{lll}
\hline\noalign{\smallskip}
Final results & $B_{s}\rightarrow\mu^{+}\mu^{-}$  barrel &  $B_{0}\rightarrow\mu^{+}\mu^{-}$  barrel \\
\noalign{\smallskip}\hline\noalign{\smallskip}
$\epsilon_{tot}$ & $0.0036\pm 0.0001$ & $0.0035\pm 0.0001$ \\ 
$N_{bkg}^{obs}$ (in SB) & \multicolumn{2}{c}{3} \\
$N_{bkg}^{exp}$ & $0.60\pm 0.35$ & $0.40\pm 0.23$  \\
$N_{peak}^{exp}$ & $0.071\pm 0.020$ & $0.245\pm 0.056$ \\
$N_{obs}$ & 2 & 0 \\
\noalign{\smallskip}\hline\noalign{\smallskip}

Final results & $B_{s}\rightarrow\mu^{+}\mu^{-}$  endcap &  $B_{0}\rightarrow\mu^{+}\mu^{-}$  endcap \\
\noalign{\smallskip}\hline\noalign{\smallskip}
$\epsilon_{tot}$ & $0.0021\pm 0.0001$ & $0.0019\pm 0.0001$ \\
$N_{bkg}^{obs}$ (in SB) & \multicolumn{2}{c}{4} \\
$N_{bkg}^{exp}$ & $ 0.80\pm 0.40$ & $ 0.53\pm0.27$ \\
$N_{peak}^{exp}$ & $ 0.044\pm 0.011$ & $ 0.158\pm0.039$ \\
$N_{obs}$ & 1 & 1 \\
\noalign{\smallskip}\hline
\end{tabular}
\end{table}

Including all statistical and systematic uncertainties the following upper limits with CLs method are extracted:
\begin{eqnarray}
BF(B_{s}\rightarrow\mu^{+}\mu^{-}) & < & 1.9\times 10^{-8} \mathrm{\; at\;  95\%\;  CL} \nonumber \\
BF(B_{0}\rightarrow\mu^{+}\mu^{-}) & < & 4.6\times 10^{-9} \mathrm{\; at\;  95\%\;  CL} \nonumber\\
BF(B_{s}\rightarrow\mu^{+}\mu^{-}) & < & 1.6\times 10^{-8} \mathrm{\; at\;  90\%\;  CL} \nonumber\\
BF(B_{0}\rightarrow\mu^{+}\mu^{-}) & < & 3.7\times 10^{-9} \mathrm{\; at\;  90\%\;  CL} \nonumber\\
\nonumber
\end{eqnarray}

\section{Conclusions}
\label{sec:concl}
A search for the rare decays $B_{s}\rightarrow\mu^{+}\mu^{-}$ and $B_{0}\rightarrow\mu^{+}\mu^{-}$  has been performed on a data sample of $pp$ collisions at $\sqrt{s} = 7$~TeV corresponding to an integrated luminosity of $1.14$~fb$^{-1}$ collected by CMS. The observed event yields are consistent with those expected adding background and SM signals. 
A combination of this result with the LHCb public result using $0.34$~fb$^{-1}$ has been performed \cite{BsCMSAndLHCB}. The combination results in an upper limit on the branching fraction of  $BF(B_{s}\rightarrow\mu^{+}\mu^{-}) < 1.1 \times 10^{-8}$ at 95\% CL. 

\end{document}